\documentclass[sigconf,screen,nonacm]{acmart}

\AtBeginDocument{%
  }

\hypersetup{
    colorlinks=true,
    linkcolor=blue,
    citecolor=magenta,
    urlcolor=teal
}

\usepackage[table]{xcolor}
\definecolor{lightgray}{gray}{0.9}

\begin{document}

\title{Measuring Human-AI Alignment as an Early Indicator for Effective Collaboration in Hybrid SE Teams \textit{(Position Paper)}}

\author{Andreas Metzger}
\email{andreas.metzger@paluno.uni-due.de}
\orcid{0000-0002-4808-8297}
\affiliation{%
  \institution{paluno (Ruhr Institute for Software Technology), University of Duisburg-Essen}
  \country{Germany}
}

\renewcommand{\shortauthors}{Metzger}

\settopmatter{printacmref=false}

\maketitle

\textit{\textbf{Motivation:}}
SE teams undergo a fundamental transformation.
AI agents, i.e., autonomous systems capable of decomposing goals, executing tasks, and collaborating with humans, are increasingly embedded into SE processes ~\cite{RoychoudhuryPPR26,OtoumE26}.
The SE community has acknowledged the rise of \emph{AI teammates}~\cite{TOSEM2026,AbrahaoGPST25} and that the evolution towards \emph{hybrid} SE teams leads to an important engineering challenge: \textbf{How to build AI agents that effectively \textit{collaborate} with humans~\cite{RoychoudhuryPPR26}?}
Such Human-AI \textit{collaboration} constitutes a significant next step from human-machine \textit{interaction}.

On the one hand, it requires finding the right scope and granularity for AI teammates~\cite{BPM2026}.
Assigning canonical SE roles (such as \emph{team member} in agile processes) to an AI agent leads to large and complex teammates that are difficult to build, test, maintain, and delegate tasks to.
Conversely, defining fine-grained, single-task agents may increase coordination overhead. 
On the other hand, not all SE tasks may be relevant in every project, and those that are may require concrete adaptations. 
As examples, formal verification may be mandatory for a safety-critical system but irrelevant for a simple utility tool; complex business logic may demand property-based tests beyond standard unit testing~\cite{InfoSys2026}.

A key prerequisite for effective collaboration~\cite{RoychoudhuryPPR26} is \textit{alignment}~\cite{GoyalCT24} -- the degree to which an AI teammate’s  granularity, its potential actions and outputs as well as constraints faithfully map to human expectations and project constraints. 
When AI teammates do not align with these expectations, they may violate process norms (which may be elicited as described in~\cite{InfoSys2026}), introduce technical debt, and force developers into exhaustive review cycles that ultimately negate the very benefits of autonomy. 
Building effective human-AI teams thus means rigorously measuring the AI teammates' adherence to human expectations~\cite{BPM2026}. 
This leads us to the research question: \textbf{How can we systematically measure human-AI \textit{alignment} in hybrid SE teams?}

Metrics for assessing hybrid teams in general exist.
Yet, most focus on measuring collaboration~\cite{A2SE2026,DellAnnaMYDGJOY26,RaufSLW25,Fragiadakis24} and not alignment~\cite{GoyalCT24}.
Measuring collaboration requires execution of the process and observing actual team dynamics and performance~\cite{DellAnnaMYDGJOY26,RaufSLW25,Fragiadakis24}.
In contrast, alignment can be measured statically, e.g., using  agent specifications, prompts, or constraints~\cite{GoyalCT24}, which facilitates early quality assurance of AI agents.

We performed a rapid literature review~\cite{Cartaxo0S18}, which indicates that only few papers consider SE-specific aspects when defining alignment metrics.
The ones that exist remain on a high level and lack an operational metric definition~\cite{RaufSLW25,Priya26}.


\vspace{1em}
\noindent\textbf{\textit{Approach:}}
Our proposed research agenda aims to deliver an SE-specific alignment metric suite, a metric validation methodology and a static measurement tool.

\textbf{1) Metric Suite:}
We will incorporate, align, adapt and integrate principles and metrics  from  the following research areas.

\textit{Software Architecture:}
 We will explore in how far classical metrics for assessing architecture quality (such as coupling and cohesion) help to assess agent granularity, which -- as motivated above -- has an impact on effective collaboration. 
 
\textit{Organizational Psychology:} The IMO (Input-Mediator-Output) framework is a classic  model used to understand how teams function and why some are more effective than others~\cite{DellAnnaMYDGJOY26}.
Here metrics, such as "capabilities and abilities" define the capacities the team collectively possesses~\cite{WilkinsSFVFW24} and thus may serve as starting point for SE-specific metrics.

\textit{HCI:} 
HCI research provides a set of assessment criteria for AI-human alignment~\cite{GoyalCT24}, which may serve as a basis for SE-specific metrics.
These HCI criteria include "Knowledge Schema" (clearly defined what specific input an agent needs), "Autonomy" (clear when the agent is allowed to act autonomously and when it has to consult a human), "Operational" (tasks and responsibilities of agent are clearly identifiable), "Accountability" (isolating, understanding, and debugging an agent error is easy), and "Human Engagement" (clear understanding when, how, and why an agent shall involve a human).

\textbf{2) Validation Methodology:}
We will empirically evaluate our alignment metrics to ensure that a metric actually measures the underlying concept it claims to represent and also to ascertain that the indeed can serve as an early indicator for effective collaboration.
We will also check inter-rater reliability and reproducibility of the static metric scoring.

One element of our validation methodology will leverage business process monitoring~\cite{CiccioM22,FrancescomarinoG22,InfoSys2023,InfoSys2026} to help track whether the statically measured alignment indeed leads to effective collaboration observed at run time and whether the measured alignment indeed helps to prevent deviations.

We will cover a broad range of evaluation entities.
Starting from agent specifications derived from canonical SE roles~\cite{HeTL25}, via project-specific roles derived from actual process logs~\cite{BPM2026}, up to micro-roles assigning only one single task to an agent.

\textbf{3) Assessment Tool: }
To statically assess alignment metrics, we will explore declarative business process modeling languages (BPM), which are well suited to capture constraints, process norms, and expected outcomes without rigidly dictating execution paths~\cite{CiccioM22,BPM2026}. 
This allows formalizing alignment metrics and statically evaluating an agent's proposed actions and in turn facilitates conformance-checking against constraints expressed in declarative models~\cite{CascianiALMS26}.
One challenge to address is how to translate the aforementioned HCI/IMO-style metrics into declarative-model-checkable constraints.




\newpage

\vspace{-1em}

\begin{table}[htbp]
\centering
\small
\caption{Experience (People, listed alphabetically)}
\rowcolors{3}{white}{lightgray}
\begin{tabular}{l c c c c c}
\toprule
\textbf{ } & \textbf{Agentic} & \textbf{Align-} & \textbf{} & \textbf{Proc-} & \textbf{Static}\\
\textbf{Person} & \textbf{SE} & \textbf{ment} & \textbf{Collab.} &\textbf{esses} & \textbf{Ana.} \\
\midrule

\href{https://scholar.google.de/citations?hl=de&user=UXe9gpwAAAAJ&view_op=list_works&sortby=pubdate}{Davide Dell'Anna} & \cite{DellAnnaMYDGJOY26} & \cite{DellAnnaMYDGJOY26} & \\

\href{https://scholar.google.de/citations?hl=de&user=ZR0Bo_QAAAAJ&view_op=list_works&sortby=pubdate}{Saimir Bala} & \cite{BPM2026} &  & \cite{BPM2026} & \cite{VavpoticBMH22} & \\

\href{https://scholar.google.de/citations?hl=de&user=nabFIW4AAAAJ&view_op=list_works&sortby=pubdate}{Matteo Esposito} & & & \cite{EspositoT26,SuEsposito26}\\

\href{https://scholar.google.de/citations?hl=de&user=8jw-zScAAAAJ&view_op=list_works&sortby=pubdate}{George Fragiadakis} & & & \cite{Fragiadakis24} & & \\

\href{https://scholar.google.de/citations?hl=de&user=vwfw0lwAAAAJ&view_op=list_works&sortby=pubdate}{Andreas Metzger} & \cite{TOSEM2026,InfoSys2026,BPM2026}  & \cite{BPM2026} &  & \cite{InfoSys2026,BPM2026} & \\

\href{https://scholar.google.de/citations?hl=de&user=8zZvFawAAAAJ&view_op=list_works&sortby=pubdate}{Andrea Marrella} & \cite{InfoSys2026}  &  & \cite{AcitelliAMM25} & \cite{InfoSys2026} & \cite{CascianiALMS26} \\


\href{https://scholar.google.de/citations?hl=de&user=SGgvbfMAAAAJ&view_op=list_works&sortby=pubdate}{Stefanie Rinderle-Ma} & 
\cite{InfoSys2026}  &  & & \cite{InfoSys2026} &  \\

\href{https://scholar.google.de/citations?hl=de&user=dK9g6a4AAAAJ&view_op=list_works&sortby=pubdate}{Klaus Schmid} & & & \cite{GroplerKJDSDTNMTVALZZ25} & & \\
\bottomrule

\end{tabular}
\end{table}

\bibliographystyle{ACM-Reference-Format-num}
\bibliography{refs,refs-am}

@article{InfoSys2026,
  author       = {Diego Calvanese and
                  Angelo Casciani and
                  Giuseppe De Giacomo and
                  Marlon Dumas and
                  Fabiana Fournier and
                  Timotheus Kampik and
                  Emanuele La Malfa and
                  Lior Limonad and
                  Andrea Marrella and
                  Andreas Metzger and
                  Marco Montali and
                  Daniel Amyot and
                  Peter Fettke and
                  Artem Polyvyanyy and
                  Stefanie Rinderle{-}Ma and
                  Sebastian Sardi{\~{n}}a and
                  Niek Tax and
                  Barbara Weber},
  title        = {Agentic Business Process Management: {A} research manifesto},
  journal      = {Inf. Syst.},
  volume       = {140},
  pages        = {102738},
  year         = {2026},
  url          = {https://doi.org/10.1016/j.is.2026.102738},
  doi          = {10.1016/J.IS.2026.102738},
  bibsource    = {dblp computer science bibliography, https://dblp.org}
}

@article{InfoSys2023,
  author       = {Andreas Metzger and
                  Tristan Kley and
                  Aristide Rothweiler and
                  Klaus Pohl},
  title        = {Automatically reconciling the trade-off between prediction accuracy
                  and earliness in prescriptive business process monitoring},
  journal      = {Inf. Syst.},
  volume       = {118},
  pages        = {102254},
  year         = {2023},
  url          = {https://doi.org/10.1016/j.is.2023.102254},
  doi          = {10.1016/J.IS.2023.102254},
  bibsource    = {dblp computer science bibliography, https://dblp.org}
}

@inproceedings{BPM2026,
      title={Using Process Mining to Generate AI Agents from Software Engineering Process Records}, 
      author={Saimir Bala and Fabiana Fournier and Lior Limonad and Andreas Metzger},
      year={2026},
      booktitle = {24th International Conference on Business Process Management (BPM 2026), Toronto, CA, Sept. 2026}, 
      url={https://arxiv.org/abs/2607.04948}, 
}

@article{TOSEM2026,
author = {Amalfitano, Domenico and Metzger, Andreas and Autili, Marco and Fulcini, Tommaso and Hey, Tobias and Keim, Jan and Pelliccione, Patrizio and Scotti, Vincenzo and Koziolek, Anne and Mirandola, Raffaela and Vogelsang, Andreas},
title = {A Research Roadmap for Augmenting Software Engineering Processes and Software Products with Generative AI},
year = {2026},
publisher = {Association for Computing Machinery},
address = {New York, NY, USA},
issn = {1049-331X},
url = {https://doi.org/10.1145/3788879},
doi = {10.1145/3788879},
journal = {ACM Trans. Softw. Eng. Methodol.},
month = jan,
}

@article{OtoumE26,
  author       = {Nesreen A. Otoum and
                  Nuha El{-}Khalili},
  title        = {Methods and Techniques of Agentic Software Engineering: {A} Systematic
                  Literature Review},
  journal      = {{IEEE} Access},
  volume       = {14},
  pages        = {7443--7465},
  year         = {2026},
  url          = {https://doi.org/10.1109/ACCESS.2026.3652325},
  doi          = {10.1109/ACCESS.2026.3652325},
  bibsource    = {dblp computer science bibliography, https://dblp.org}
}

@article{HeTL25,
  author       = {Junda He and
                  Christoph Treude and
                  David Lo},
  title        = {LLM-Based Multi-Agent Systems for Software Engineering: Literature
                  Review, Vision, and the Road Ahead},
  journal      = {{ACM} Trans. Softw. Eng. Methodol.},
  volume       = {34},
  number       = {5},
  pages        = {124:1--124:30},
  year         = {2025},
  url          = {https://doi.org/10.1145/3712003},
  doi          = {10.1145/3712003},
  bibsource    = {dblp computer science bibliography, https://dblp.org}
}

@inproceedings{GoyalCT24,
  author       = {Nitesh Goyal and
                  Minsuk Chang and
                  Michael Terry},
  editor       = {Florian 'Floyd' Mueller and
                  Penny Kyburz and
                  Julie R. Williamson and
                  Corina Sas},
  title        = {Designing for Human-Agent Alignment: Understanding what humans want
                  from their agents},
  booktitle    = {{CHI} Conference on Human Factors in Computing
                  Systems},
  pages        = {106:1--106:6},
  publisher    = {{ACM}},
  year         = {2024},
  url          = {https://doi.org/10.1145/3613905.3650948},
  doi          = {10.1145/3613905.3650948},
  bibsource    = {dblp computer science bibliography, https://dblp.org}
}

@article{RoychoudhuryPPR26,
  author       = {Abhik Roychoudhury and
                  Corina S. Pasareanu and
                  Michael Pradel and
                  Baishakhi Ray},
  title        = {Agentic {AI} Software Engineers: Programming with Trust},
  journal      = {Commun. {ACM}},
  volume       = {69},
  number       = {5},
  pages        = {56--58},
  year         = {2026},
  url          = {https://doi.org/10.1145/3769314},
  doi          = {10.1145/3769314},
  bibsource    = {dblp computer science bibliography, https://dblp.org}
}

@article{Fragiadakis24,
  author       = {George Fragiadakis and
                  Christos Diou and
                  George Kousiouris and
                  Mara Nikolaidou},
  title        = {Evaluating Human-AI Collaboration: {A} Review and Methodological Framework},
  journal      = {CoRR},
  volume       = {abs/2407.19098},
  year         = {2024},
  url          = {https://doi.org/10.48550/arXiv.2407.19098},
  doi          = {10.48550/ARXIV.2407.19098},
  eprinttype   = {arXiv},
  eprint       = {2407.19098},
  bibsource    = {dblp computer science bibliography, https://dblp.org}
}

@article{DellAnnaMYDGJOY26,
  author       = {Davide Dell'Anna and
                  Pradeep K. Murukannaiah and
                  Mireia Yurrita and
                  Bernd Dudzik and
                  Davide Grossi and
                  Catholijn M. Jonker and
                  Catharine Oertel and
                  Pinar Yolum},
  title        = {From human teams to hybrid intelligence teams: identifying, characterizing,
                  and evaluating foundational quality attributes},
  journal      = {Auton. Agents Multi Agent Syst.},
  volume       = {40},
  number       = {1},
  pages        = {10},
  year         = {2026},
  url          = {https://doi.org/10.1007/s10458-025-09730-8},
  doi          = {10.1007/S10458-025-09730-8},
  bibsource    = {dblp computer science bibliography, https://dblp.org}
}

@article{WilkinsSFVFW24,
  author       = {Jay Wilkins and
                  David A. Sparrow and
                  Caitlan A. Fealing and
                  Brian D. Vickers and
                  Kristina A. Ferguson and
                  Heather Wojton},
  title        = {A team-centric metric framework for testing and evaluation of human-machine
                  teams},
  journal      = {Syst. Eng.},
  volume       = {27},
  number       = {3},
  pages        = {466--484},
  year         = {2024},
  url          = {https://doi.org/10.1002/sys.21730},
  doi          = {10.1002/SYS.21730},
  bibsource    = {dblp computer science bibliography, https://dblp.org}
}

@incollection{Priya26,
  title={Building AI-Augmented Teams},
  author={Priya, A},
  booktitle={Generative AI in Software Development},
  pages={185--207},
  year={2026},
  publisher={Productivity Press},
  url ={https://www.taylorfrancis.com/chapters/edit/10.4324/9781003729068-12/building-ai-augmented-teams-priya}
}

@inproceedings{RaufSLW25,
  author       = {Irum Rauf and
                  Helen Sharp and
                  Tamara Lopez and
                  Michel Wermelinger},
  title        = {Human-Machine Teaming and Team Effectiveness in {AI} Tools for Software
                  Engineering},
  booktitle    = {18th {IEEE/ACM} International Conference on Cooperative and Human
                  Aspects of Software Engineering, CHASE@ICSE 2025, Ottawa, ON, Canada,
                  April 27-28, 2025},
  pages        = {75--80},
  publisher    = {{IEEE}},
  year         = {2025},
  url          = {https://doi.org/10.1109/CHASE66643.2025.00017},
  doi          = {10.1109/CHASE66643.2025.00017},
  bibsource    = {dblp computer science bibliography, https://dblp.org}
}

@article{AbrahaoGPST25,
  author       = {Silvia Abrah{\~{a}}o and
                  John C. Grundy and
                  Mauro Pezz{\`{e}} and
                  Margaret{-}Anne D. Storey and
                  Damian A. Tamburri},
  title        = {Software Engineering by and for Humans in an {AI} Era},
  journal      = {{ACM} Trans. Softw. Eng. Methodol.},
  volume       = {34},
  number       = {5},
  pages        = {129:1--129:46},
  year         = {2025},
  url          = {https://doi.org/10.1145/3715111},
  doi          = {10.1145/3715111},
  bibsource    = {dblp computer science bibliography, https://dblp.org}
}

\end{document}